\documentclass[aps,showpacs,preprintnumbers,amsmath,amssymb]{revtex4}

\usepackage{graphicx}
\usepackage{dcolumn}
\usepackage{bm}
\usepackage{hyperref}

\newcommand{\beq}{\begin{equation}}
\newcommand{\eeq}{\end{equation}}
\newcommand{\beqa}{\begin{eqnarray}}
\newcommand{\eeqa}{\end{eqnarray}}
\newcommand{\beqar}{\begin{eqnarray*}}
\newcommand{\eeqar}{\end{eqnarray*}}
\newcommand{\bra}[1]{\mbox{$\left\langle{#1}\right|$}}
\newcommand{\ket}[1]{\mbox{$\left|{#1}\right\rangle$}}

\def\I{{\rm i}}

\def\e{{\rm e}}
\def\Tr{{\rm Tr}}

\newcounter{saveeqn}
\newcommand{\alpheqn}{\setcounter{saveeqn}{\value{equation}}%
\stepcounter{saveeqn}\setcounter{equation}{0}%
\renewcommand{\theequation}{\mbox{\arabic{saveeqn}\alph{equation}}}}
\newcommand{\reseteqn}{\setcounter{equation}{\value{saveeqn}}%
\renewcommand{\theequation}{\arabic{equation}}}
\def\beql{\alpheqn \beqa}
\def\eeql{\eeqa \reseteqn}

\begin{document}
\title{Proposal of Quantum Simulation
of Pairing Model\\ on an NMR Quantum Computer}
\author{An Min Wang}
\affiliation{Department of Modern Physics, University of Science
and Technology of China, Hefei, 230026, P.R.China}
\author{Xiaodong Yang}
\affiliation{Department of Modern Physics, University of Science
and Technology of China, Hefei, 230026, P.R.China}

\begin{abstract}
We give out a proposal of quantum simulation of pairing model on
an NMR quantum computer. In our proposal, we choose an appropriate
initial state which can be easily prepared in experiment. Making
use of feature of NMR measure and the technology of the second
(discrete) Fourier transformation, our theoretical scheme can
obtain the spectrum of paring model in principle. We concretely
discuss the case in the concerned subspaces of pairing model and
then, as an example, give out a simple initial state to get the
gap of two the lowest energy levels in the given subspace. The
quantum simulation to get more differences of energy levels is
able to be discussed similarly.

\end{abstract}

 \maketitle


Simulating a real physical system by a quantum computer (QC) was
originally conjectured by Feymann \cite{FeynmanQS}. Later, this
idea was confirmed by Llody in a two-state array \cite{LloydQS}
and a general scheme for the quantum simulation was presented
\cite{QS}.

NMR quantum computer is one of successful realizations of quantum
computer so far \cite{NMRQCReview,Jones}. Moreover, a four-level
truncted oscillator \cite{QS}, a three-spin effective Hamiltonian
\cite{Tseng3spin} and the migration of excitation in an
eight-state quantum system \cite{Khitrin} have been simulated. In
recent, \textsl{L.-A. Wu} \textsl{et al}. \cite{simulation}
reported an NMR experiment scheme performing a polynomial-time
simulation of pairing models. Moreover, new works are coming forth
continually \cite{Somma,Negrevergne}. In this letter, we give out
a proposal of quantum simulation of pairing model on an NMR
quantum computer. Two main features of our proposal are: (1) the
choose of an appropriate initial state which can be easily
prepared in experiment; (2) the using the second (discrete)
Fourier transformation which can obtain the spectrum of paring
model. We concretely discuss the case in the concerned subspaces
of pairing model and then, as an example, give out a simple
initial state to get the gap of two the lowest energy levels in
the given subspace. The quantum simulation to get more differences
of energy levels is able to be discussed similarly. In order to
know what is the theoretical foundation of our scheme and whether
our scheme can arrive at the needed precision, we also carry out
some relevant research, see our preprints\cite{Oursd, Ournsd}. In
addition, after proposing our theoretical scheme we also finished
its simple experimental implement \cite{Ourqc}.

Let us start from the spin-analogy of paring model Hamiltonian
\cite{simulation,solid} \beq \label{hspinorig}
H_p=\sum_{m=1}^{N}\frac{\epsilon_{m}}{2}\sigma_{z}^{(m)}
-\frac{V}{2}\sum_{m<l=1}^{N}\left(\sigma_{x}^{(m)}\sigma_{x}^{(l)}%
+\sigma_{y}^{(m)}\sigma_{y}^{(l)}\right)\eeq

 {\em We first generally prove that if an appropriate working
initial state is chosen rightly, the differences of energy levels
of $H_p$ can be obtained by using of two times Fourier
transformations.}

Without loss of generality, a general working initial state can be
written as \beq \label{inis} \ket{\psi_{\rm ini}}=\sum_i a_i
\ket{v_p^i}=\sum_{i,j}a_ib_{ij}\ket{j}=\sum_{i,j}a_ib_{ij}\ket{\phi_{\rm
nmr}^j}\eeq where $\ket{v_p^i}=\sum_j b_{ij}\ket{j}$ are the
$H_p$'s eigenvectors with the corresponding eigenvalues $E_p^{i}$,
and $\ket{j}$ are the standard spin basis. Here, we have used the
fact that the eigenvectors $\ket{\phi_{\rm nmr}^j}$ of $H_{\rm
nmr}$ with eigenvalues $E_{\rm nmr}^j$ are just $\ket{j}$ since
$H_{\rm nmr}$ in the laboratory is diagonal. \beq H_{\rm
nmr}=\frac{1}{2}\left(\sum_{i=1}^N\omega_0^i\sigma^i_z+\sum_{i,j=1,i<j}^N\pi
J\sigma_z^i\sigma_z^j\right)\eeq Considering the following
evaluation \beq \e^{-\I H_p \tau/\hbar}\ket{\psi_{\rm ini}}=\sum_i
a_i \e^{-\I E_p^i \tau/\hbar}\ket{v_p^i}\eeq one obtains the
density matrix \beq
\rho(\tau)=\sum_{i,i^\prime}\sum_{j,j^\prime}a_i
b_{ii^\prime}a_j^*b_{jj^\prime}^*\exp\{-\I\left(E_p^i-E_p^j\right)\tau/\hbar\}\ket{\phi_{\rm
nmr}^{i^\prime}}\bra{\phi_{\rm nmr}^{j^\prime}} \eeq

Recalling the procedure of the NMR measure, one can write the
result, that is the NMR frequency spectrum, as \beq S_{\rm
NMR}(\omega)\propto ft\left[\Tr\left(\e^{-\I H_{\rm
nmr}t/\hbar}\rho_{f}\e^{\I H_{\rm
nmr}t/\hbar}\sum_{k=1}^N\sigma^{+}_k\right)\right]\eeq where
$\sigma_i^{\pm}=(\sigma_x^i\pm\I\sigma_y^i)/2$, $\rho_f$ is the
density matrix to be measured and $ft$ means the Fourier
transformation. It must be emphasized that the Fourier
transformation is applied for the NMR measure time $t$.

Now, substituting $\rho(\tau)$ into the above equation, we have
\beq S_{\rm NMR}(\omega)\propto
\sum_{i,i^\prime}\sum_{j,j^\prime}a_i
b_{ii^\prime}a_j^*b_{jj^\prime}^*\exp\{-\I\left(E_p^i-E_p^j\right)\tau/\hbar\}
\delta(E_{\rm nmr}^{i^\prime}-E_{\rm nmr}^{j^\prime})
\Tr\left(\ket{i^\prime}\bra{j^\prime}\sum_{k=1}^N\sigma^{+}_k\right)\eeq
For an appearing peak in NMR frequency spectrum, for example
$E_{\rm nmr}^{\alpha}-E_{\rm nmr}^{\beta}$, we can know its
corresponding difference of energy levels (the values of $\alpha$
and $\beta$) based on NMR Hamiltonian and the parameters of
sample. Then, we carry out the second Fourier transformation and
are able to obtain the $H_p$'s frequency spectrum \beq
S_p(E_p)\propto \sum_{i,j}a_i
b_{i\alpha}a_j^*b_{j\beta}^*\delta(E_{p}^{i}-E_{p}^{j})\eeq In
experiment, the second (discrete) Fourier transformation can be
done from the dada that are collected by measuring the areas
(heights) of peaks $E_{\rm nmr}^{\alpha}-E_{\rm nmr}^{\beta}$ at a
series of evolution times $\tau_i$ with $H_p$. As soon as the
$E_{p}^{i}-E_{p}^{j}$ are given out and the lowest energy level is
also known, the spectrum of $H_p$ is just obtained.

However, there are still three difficulties facing on us.{\em (1)
Which state is an appropriate working initial state for our
purpose; (2) Which $E_{\rm nmr}^{\alpha}-E_{\rm nmr}^{\beta}$
peaks can appear in NMR frequency spectrum for a given initial
stste; (3) What $i$ and $j$ values correspond to the
$E_{p}^{i}-E_{p}^{j}$ peaks in $H_p$'s frequency spectrum.}

In order to solve them, we need some mathematical and physical
preparations.

Firstly, we should derive out the obvious expression
$\Tr\left(\ket{i^\prime}\bra{j^\prime}\sum_{k=1}^N\sigma^{+}_k\right)$.
Note the fact that the spin space can be divided into the
different subspaces which correspond to the different numbers of
spin-up states, that is $S_{\rm spin}^{(N)}=S_{0}^{(N)}\oplus
S_{1}^{(N)}\oplus S_{2}^{(N)}\oplus\cdots\oplus S_{N}^{(N)}$,
where the subspace $n$, $i.e$ $S_n^{(N)}$, is a subspace with $n$
spin-up states $\ket{0}$. It is clear that these subspaces contain
the following basis \beql S_0^{(N)}&=&\{\ket{2^N}\},\qquad
S_N^{(N)}=\{\ket{1}\}\\
S_n^{(N)}&=&\left\{\ket{s^{(N)}_{i_1i_2\cdots
i_n}}=\ket{2^N-\displaystyle\sum_{a=1}^n 2^{(N-i_a)}}, i_a\neq
i_b, i_a=1,2,\cdots,N\right\}\eeql \noindent Further, based on the
relations $\sigma^+\ket{0}=0$, $\sigma^+\ket{1}=\ket{0}$,
$\sigma^-\ket{0}=\ket{1}$ and $\sigma^-\ket{1}=0$, we have that
\beql \sum_{k=1}^N\sigma_k^-
\ket{1}=\sum_{k=1}^N\ket{2^{(N-k)}+1},\quad \sum_{k=1}^N\sigma_k^-
\ket{s^{(1)}_{i_1}}=\ket{2^N},\quad \sum_{k=1}^N\sigma_k^-
\ket{2^N}=0\\ \sum_{k=1}^N\sigma_k^- \ket{s_{i_1i_2\cdots
i_n}^{(N)}} =\sum_{k=
i_1,\cdots,i_n}\ket{2^N-\displaystyle\sum_{a=1}^n
2^{(N-i_a)}+2^{N-k}}\qquad\eeql In terms of these equations, it is
easy to get \beqa
\Tr\left(\ket{i}\bra{j}\sum_{k=1}^N\sigma^+_k\right)
&=&\sum_{k=1}^N\delta_{i,2^{(N-k)}+1}\delta_{j1}
+\delta_{i,2N}\sum_{k=1}^N\delta_{j,2^N-2^{(N-k)}}\nonumber\\
& &
+\sum_{k=i_1,i_2,\cdots,i_n}\delta_{i,j+2^{(N-k)}}\delta_{j,2^N-\sum_{a=1}^n
2^{(N-i_a)}} \eeqa where in the last term $n$ with possible values
$2,3,\cdots,N-1$, $i_a\neq i_b$ and $i_a=1,2,\cdots,N$ for any
$a=1,2,\cdots,n$.

Secondly, we also need to analyze the structure of the
eigenvectors of $H_p$. From eq.(\ref{hspinorig}) and the relations
$\left(\sigma_x^{(m)}\sigma_x^{(l)}+\sigma_y^{(m)}\sigma_y^{(l)}\right)
=\left(\sigma_x^{(m)}+\I\sigma_y^{(m)}\right)
\left(\sigma_x^{(l)}-\I\sigma_y^{(l)}\right), (m\neq l)$, they
follow that $\ket{1}$ and $\ket{2^N}$ must be $H_p$'s two
eigenvectors respectively corresponding to the maximum and the
minimum eigenvalues, which are denoted respectively by
$\ket{v_p^1}$ and $\ket{v_p^{2^N}}$. Moreover, if the arbitrary
basis $\ket{s_{i_1\cdots i_n}^{(N)}}$ belongs to $S_n^{(N)}$, then
$H_p\ket{s_{i_1\cdots i_n}^{(N)}}$ also belongs to $S_n^{(N)}$
because that $\sigma^+$ and $\sigma^-$ appear in pairs or do not
appear in the various terms of $H_p$. This implies that
$\bra{s_{i_1\cdots i_{m}}^{(N)}}H_p\ket{s_{i_1\cdots
i_n}^{(N)}}=0, (m\neq n; m,n=1,2,\cdots, N-1)$. Therefore
\begin{equation}
H_{p}^{(N)}=H_{sub0}^{(N)}\oplus H_{sub1}^{(N)}\oplus
H_{sub2}^{(N)}\oplus\cdots\oplus
H_{subN}^{(N)} \label{dirsum1}%
\end{equation}
So we can denote the others eigenvectors of $H_p$ as
$\ket{v_p^{i^{(n)}}}=\sum_{j^{(n)}}b_{i^{(n)}j^{(n)}}\ket{j^{(n)}}\in
S_n^{(N)}$ with the corresponding eigenvalues $E_p^{i^{(n)}}$, and
$i^{(n)},j^{(n)}$ only take the sequence number of spin basis
belonging to the subspace $S_n^{(N)}$, for example
$i^{(1)}(k)=2^N-2^{(N-k)}, (k=1,2,\cdots, N)$. In other words,
$\ket{i^{(n)}},\ket{j^{(n)}}\in S_n^{(N)}$.

Now let us solve the difficulty one. If we only concern the
spectrum $E_p^{i^{(1)}}$ and $E_p^{i^{(N-1)}}$ in the subspace 1
and subspace $N-1$ respectively, we should choose such a working
initial state that it does not include any $\ket{v_p^{i^{(m)}}},
(m=2,3,\cdots N-2)$. Obviously, the appearing NMR frequency
spectrum will be simplified as at most $2N$ peaks. \beqa S_{\rm
NMR}(\omega)&\propto& \sum_{k}^N{\sum_{i,j}}^\prime \left[a_i
b_{i,2^{(N-k)}+1}a_j^*b_{j1}^*\exp\{-\frac{\I}{\hbar}\left(E_p^i-E_p^j\right)\tau\}
\delta(E_{\rm nmr}^{2^{(N-k)}+1}-E_{\rm nmr}^{1}) \right.\nonumber\\
& &\left.+a_i
b_{i,2^{N}}a_j^*b_{j,2^{N}-2^{(N-k)}}^*\exp\{-\frac{\I}{\hbar}\left(E_p^i-E_p^j\right)\tau\}
\delta(E_{\rm nmr}^{2^N}-E_{\rm
nmr}^{2^{N}-2^{(N-k)}})\right]\eeqa where in the summation
$\displaystyle{\sum_{i,j}}^\prime$, $i,j$ only take over
$1,2^{(N-k)}+1,2^N-2^{(N-k)},2^N; (k=1,2,\cdots, N)$. In this
simplified case, the NMR spectrum only include the differences of
energy levels in the subspaces 0, 1, $N-1$ and $N$. In particular,
if the working initial state is taken as
$c\ket{v_p^{i^{(1)}}}+d\ket{v_p^{j^{(1)}}}$, then
$E_p^{i^{(1)}}-E_p^{j^{(1)}}$ can be obtained from the second
Fourie transformation unless no any peaks appears in the NMR
frequency spectrum.

Then, let us solve the difficulty two, that is, how to guarantee
the NMR frequency spectrum has those needed peaks. The key matter
is that we have to take an appropriate working initial state
including the kets $\ket{1}$, $\ket{2^{(N-k)}+1}(\in
S^{(N)}_{(N-1)})$ or(/and) $\ket{2^N-2^{(N-k)}}(\in S^{(N)}_1)$,
$\ket{2^N}$, that is $\ket{\psi_{\rm
ini}}=a_1\ket{1}+\sum_{i^{(N-1)}}a_{i^{(N-1)}}\ket{v_p^{i^{(N-1)}}}$
or $\ket{\psi_{\rm
ini}}=a_{2^N}\ket{2^N}+\sum_{i^{(1)}}a_{i^{(1)}}\ket{v_p^{i^{(1)}}}$.
Respectively, for the two cases we have \beqa S_{\rm
NMR}(\omega)\!\!\!&\propto&\!\!\! \sum_{k}^N\sum_{i^{(N-1)}}
a_{i^{(N-1)}}
b_{i^{(N-1)},2^{(N-k)}+1}a_1^*b_{11}^*\exp\{-\frac{\I}{\hbar}\left(E_p^{i^{\!(N-1)\!}}\!\!-E_p^1\right)\tau\}
\delta(E_{\rm nmr}^{2^{(N-k)}+1}-E_{\rm nmr}^{1})\\
S_{\rm NMR}(\omega)\!\!\!&\propto&\!\!\! \sum_{k}^N\sum_{j^{(1)}}
a_{2^N}
b_{2^N,2^N}a_{j^{(1)}}^*b_{j^{(1)},2^N-2^{(N-k)}}^*\exp\{-\frac{\I}{\hbar}\left(E_p^{2^N}-E_p^{j^{(1)}}\right)\tau\}
\delta(E_{\rm nmr}^{2^N}-E_{\rm nmr}^{j^{(1)}})\eeqa \noindent
Since $a_{i^{(N-1)}} b_{i^{(N-1)},2^{(N-k)}+1}a_1^*b_{11}^*$ or
$a_{2^N} b_{2^N,2^N}a_{j^{(1)}}^*b_{j^{(1)},2^N-2^{(N-k)}}^*$ are
all not zero, NMR spectrum must have some peaks appearing.

The last difficulty is easy to be solved. In fact, setting
$\ket{\psi_{\rm
ini}}=a_{2^N}\ket{2^N}+a_{2^N-2^{(N-K)}}\ket{v_p^{2^N-2^{(N-K)}}}$,
we can read $E_p^{2^N-2^{(N-K)}}-E_p^{2^N}$ from $H_p$'s spectrum
and then obtain $E_p^{2^N-2^{(N-K)}}$ in terms of
$E_p^{2^N}=-\sum_{m=1}^N\epsilon_m/2$. If setting $\ket{\psi_{\rm
ini}}=a_{2^N}\ket{2^N}+a_{2^N-2^{(N-k_1)}}\ket{v_p^{2^N-2^{(N-k_1)}}}
+a_{2^N-2^{(N-k_2)}}\ket{v_p^{2^N-2^{(N-k_2)}}},(k_1\neq k_2)$, we
can read $E_p^{2^N-2^{(N-k_1)}}-E_p^{2^N}$ and
$E_p^{2^N-2^{2^N-(N-k_2)}}-E_p^{2^N}$, and then obtain
$E_p^{2^N-2^{(N-k_1)}}-E_p^{2^N-2^{(N-k_2)}}$ (set
$E_p^{2^N-2^{(N-k_1)}}\geq E_p^{2^N-2^{(N-k_2)}}$ if $k_2>k_1$).
Therefore, in principle, we always can find out the energy levels
and their differences in the subspace 1 of $H_p$. Likewise, one
ought to able to obtain the other energy levels in the other
subspaces by the different chooses of the working initial states.

It must be emphasized that it is interesting what is the physical
meaning of $E_p^{2^N-2^{(N-k_1)}}-E_p^{2^N-2^{(N-k_2)}}$ in theory
and what is an appropriate working initial state to obtain
$E_p^{2^N-2^{(N-k_1)}}-E_p^{2^N-2^{(N-k_2)}}$ in experiment.

Actually, by the submatrix diagonalization of spin-analogy of
pairing model \cite{Oursd} and the numerical calculation
\cite{Ournsd}, we have found that the relation when $N$ is large
enough
\begin{equation}
(\xi_{2^N-2}^{2}+\Delta^{2})^{1/2}-(\xi_{2^N-1}^{2}+\Delta^{2})^{1/2}\approx
E_p^{2^N-2}-E_p^{2^N-1}
\label{gapeqnew}%
\end{equation}
\noindent where $E_p^{2^N-2}-E_p^{2^N-1}$ is the difference of two
the lowest energy levels in the subspace 1 of $H_p$, $\Delta$ is
the solution of the energy gap equation \cite{taylor,li}
\begin{equation}
1=\frac{1}{2}V\sum_{m}\frac{1}{\sqrt{\xi_{m}^{2}+\Delta^{2}}} \label{gapeqold}%
\end{equation}
\noindent while $\xi_m$ comes from
\begin{equation}
\widetilde{H}_{\mathrm{BCS}}=\varepsilon_{s}+\frac{1}{2}\sum_{m=1}^{N}(\xi
_{m}^{2}+\Delta^{2})^{1/2}(\gamma_{m}^{\dagger}\gamma_{m}+\gamma_{-m}%
^{\dagger}\gamma_{-m}) \label{rhbcs}%
\end{equation}\noindent
Here $\gamma_{m}^{\dagger}$, $\gamma_{m}$ are the quasiparticle
creation and annihilation operators; $\varepsilon_{s}$ is the
ground state energy of superconducting system. Hamiltonian
(\ref{rhbcs}) was obtained by use of Bogoliubov transformation and
mean field approximation\cite{taylor,li} to the BCS Hamiltonian
$H_{\mathrm{BCS}}=\sum_{m=1}^{N}(\varepsilon_{m}-\varepsilon_{F})
(n_{m}+n_{-m})/2-V\sum_{m,l=1}^{N}c_{m}^{\dagger}c_{-m}^{\dagger}c_{-l}c_{l}
\label{bcsh}$. This result is consist with the known conclusion.
Thus, we know the physical meaning of
$E_p^{2^N-2^{(N-k_1)}}-E_p^{2^N-2^{(N-k_2)}}$ in theory.

In order to see what working initial state should be chosen to
obtain $E_p^{2^N-2^{(N-k_1)}}-E_p^{2^N-2^{(N-k_2)}}$ in
experiment, let us consider an example when $k_2=1,k_1=2$. It is
easy to prove that \beqa
H_p\ket{\overline{W}}&=&-\left[\frac{1}{2}\sum_{m=1}^N\epsilon_m+(N-1)V\right]
\ket{\overline{W}}+
\frac{1}{\sqrt{N}}\sum_{m=1}^M\epsilon_m\ket{11\cdots
1\underbrace{0}_m
1\cdots 1}\\
H_p\ket{\overline{u}_{ij}}&=&
-\left[\frac{1}{2}\sum_{m=1}^N\epsilon_m-\epsilon_j-V\right]
\ket{\overline{u}_{ij}}-(\epsilon_j-\epsilon_i)\frac{1}{\sqrt{2}}\ket{11\cdots
1\underbrace{0}_i 1\cdots 1}\eeqa \noindent where we have defined
\beq
\ket{\overline{W}}=\frac{1}{\sqrt{N}}\sum_{i=1}^N\ket{11\cdots
1\underbrace{0}_i 1\cdots 1},\quad
\ket{\overline{u}_{ij}}=\frac{1}{\sqrt{2}}\left(\ket{11\cdots
1\underbrace{0}_i 1\cdots 1}-\ket{11\cdots 1\underbrace{0}_j
1\cdots 1}\right) \eeq Here, $i\neq j$ and $i,j=1,2,\cdots, N$.
Obviously, $\ket{\overline{u}_{ij}}$ are not completely
independent each other. In practice, we can fix $i=1$ and
$j=2,3,\cdots,N$ and obtain $(N-1)$ linearly independent
$\ket{\overline{u}_{ij}}$. Note that when all $\epsilon_m$ are
equal, so-called anti-W state $\ket{\overline{W}}$ corresponds to
the lowest energy level in the subspace 1. When $\epsilon_m$ are
different for the different $m$ and $\epsilon_{j+1}-\epsilon_j$ is
a very small positive parameter, we know, from the perturbation
theory, that the lowest energy level is still related with
$\ket{\overline{W}}$, and the second lowest energy ought to relate
with $\ket{\overline{u}_{12}}$. Thus, in order to make the working
initial state to contain $\ket{v_p^{(2^N-1)}}$ and
$\ket{v_p^{(2^N-2)}}$ definitely, we should take it as \beq
\label{ourinis} \ket{\psi_{\rm
ini}^{(0)}}=\frac{1}{\sqrt{3}}\left(\ket{2^N}+\ket{\overline{W}}+\ket{\overline{u}_{12}}\right)
\eeq \noindent Thus, the final $H_p$'s frequency spectrum includes
consequently the peaks $E_p^{2^N-1}-E_p^{2^N}$ and
$E_p^{2^N-2}-E_p^{2^N}$. Obviously, such an initial state is
simple and easy to prepare in NMR experiment.

In the summary, our scheme of quantum simulation of paring model
on an NMR quantum computer has four steps. (1) prepare the initial
state based on the quantum simulation purpose, for example
(\ref{ourinis}) in order to obtain $E_p^{2^N-2}-E_p^{2^N-1}$; (2)
use NMR pulse series to implement evolution $\e^{-\I
H_p\tau/\hbar}$ \cite{simulation, Ourqc} at a series of times
$\tau_i$; (3) carry out NMR measure (the first Fourier
transformation) to clollect data which are a series of amplitudes
corresponding to the evolution time $\tau_i$ by $H_p$ ; (4) make
the second (discrete) Fourier transformation for the collecting
data and then obtain $H_p$'s frequency spectrum. Newly, we have
finished the experiment of quantum simulation for the simplest
system of two qubits. More experiment detail is put in our another
paper \cite{Ourqc}.

Comparing with the known scheme \cite{simulation} and our one,
obviously, the working initial state in the former are prepared by
the process of a quasi-adiabatically evolution, but our scheme
does not need such a process and chose directly an appropriate
working initial state which can be easily prepared in experiment.
In terms of feature of NMR measure and the second (discrete)
Fourier transformation, our scheme can obtain the spectrum of
paring model in principle. It is different from the scheme in
ref.\cite{simulation} where the second Fourier transformation was
not used. In addition, because we chose an appropriate working
state, the rotation step is not needed. However, it must point out
that for every time evolution, we use the same serial of pulses as
the ref.\cite{simulation}. These features of our proposal leads
that the simulating paring model on an NMR quantum computer is
actually feasible and really complete. Moreover, we concretely
discuss the case in the subspaces 1 and $N-1$ of pairing model and
then, as an example, give out a simple initial state to get the
gap of two the lowest energy levels in the subspace 1. The quantum
simulation to get more differences of energy levels is able to be
discussed similarly. This working is on progressing.

We particularly thank Jiangfeng Du for his valuable suggestions
and indispensable supports in the experiment implement of our
scheme, and we are grateful Feng Xu, Xiaosan Ma, Niu Wanqing,
Ningbo Zhao, Hao You, Zhu Rengui and Su Xiaoqiang for helpful
discussion. This work was founded by the National Fundamental
Research Program of China with No. 2001CB309310, and partially
supported by the National Natural Science Foundation of China
under Grant No. 60173047.

\end{document}